\documentclass{appolb}
\usepackage{graphicx}
\usepackage{xcolor}
\usepackage{subfig}

\begin{document}
\title{Equation of State at High-Baryon Density\\
and Compact Stellar Objects%
\thanks{Presented at Quark Matter 2022}%
}
\author{Veronica Dexheimer
\address{Department of Physics, Kent State University, Kent, OH 44242, USA}
}
\maketitle
\begin{abstract}
In this contribution I review the connection between compact stars and high-baryon density matter, focusing on astrophysical observables for deconfinement to quark matter. I discuss modern ingredients, repositories, and constraints for the neutron-star equations of state. Finally, I draw comparisons between dense and hot matter created in neutron-star mergers and heavy-ion collisions, and the possibility of quantitatively establishing a link between them.
\end{abstract}
  
\section{Introduction to compact stars}

Compact stars are the endpoints of stellar evolution, and refer to white dwarfs, neutron stars (NSs), and black holes. The different evolution endpoints are determined by the mass of the stars in the beginning of their lives. Stars that begin their evolution more massive, move faster through the different fusion cycles and collapse more violently when they run out of fuel for fusion. The gravitational collapse comes to a halt when there is enough degenerate matter pressure to balance gravity once more. The degenerate matter pressure is provided by an electron gas in white dwarfs, and mainly by a neutron gas in NSs, with contributions (in the latter) from electrons, muons, protons, and possibly hyperons and quarks. But, more importantly, in NSs the additional compression due to stronger gravity also brings baryons close enough for the strong force to contribute significantly to hydrostatic equilibrium. This is evidenced by the fact that if NSs are modeled as a free gas of baryons, one does not reproduce any of the observed NS masses ($1-2$ M$_\odot$), generating errors of $\sim100\%$ \cite{Sagert:2005fw}. Finally, for extremely massive stars, the gravitational collapse never halts, giving birth to black holes.

Intermediate-mass stars explode as core-collapse supernovae and leave behind small NSs, of the size of a city ($\sim12$ km of radius). They span densities from $1$ g/cm$^3$ in the atmosphere to about $10^{15}$ g/cm$^3$ — corresponding to a number density of about 1 baryon per fm$^3$ — in the center. The crust extends until around nuclear saturation density $n_0\sim0.15$ fm$^{-3}$, where nuclei dissolve into bulk baryonic matter in the core. At a few times $n_0$ it becomes energetically favored for nucleons to convert into hyperons \cite{Glendenning:1982nc} or spin $3/2$ $\Delta$ baryons \cite{Prakash:1992zng} (see Fig.~1 in Ref.~\cite{Weber:2014qoa}). Eventually, all baryons deconfine into quarks, the question being, are massive NS cores dense enough for deconfinement to take place in their interiors? And if they are, can deconfinement precede the appearance of hyperons?

\section{Neutron-star equation of state}

Formally, the equation of state (EoS) is a thermodynamic equation relating state variables, usually referring to the pressure as a function of energy density. More broadly, it can include a full list of thermodynamical variables, particle composition, microscopic information, and stellar properties. Although NSs can be considered to have zero temperature (in MeV scale) and be in $\beta-$equilibrium with leptons, the matter in proto NSs or in hypermassive stars formed in NS mergers is hot and not equilibrated. As such, supernova and merger simulations require 3-dimensional EoS tables as input. These usually include temperature $T$ and charge fraction $Y_Q$, in addition to baryon number density $n_B$, as independent variables.

A complete EoS for NSs must include atmosphere and crust components connected to the core. Due to competing effects between strong and electromagnetic forces, such a connection is far from being simple due to the formation of a diversity of shapes, the so-called pasta phase. Although the pasta does not influence stellar properties such as mass and radius, it modifies, for example, the interactions of neutrinos with the nuclear medium and is important for the dynamics of supernovae \cite{Horowitz:2004yf}. Due to the attractive component of the strong force, fermionic matter at high densities (and low temperatures) is paired. This feature has also been shown to be important for the dynamics of baryonic matter in both the crust and core and, in the case of deconfined matter, for the quarks \cite{Baldo:1992kzz,Page:2010aw,Sedrakian:2004yq,Alford:2001dt,Ivanytskyi:2022oxv}.

Modern online repositories provide 1-, 2-, and 3-dimensional EoS tables for astrophysical applications. The largest example is CompOSE \cite{compose}, now offering hundreds of different EoS's, together with software to interpolate data, calculate additional quantities, and graph EoS dependencies (see full instruction manual in Ref.~\cite{Typel:2022lcx}). Another idea is to offer a Modular Unified Solver of the Equation of State (MUSES) \cite{muses}. While currently under development, in a few years MUSES will be offering customized EoSs that will be fitted and combined at will to cover any desired portion of the high-energy QCD phase diagram.

\section{Equation of state constraints}

The most important qualities for an EoS are being thermodynamical consistent and remaining causal, the latter being guaranteed by relativistic models, as long as vector interactions are not too strong \cite{Zeldovich:1961sbr}. Isospin-symmetric (effectively) zero-temperature baryonic matter can be constrained around $n_0$ from low-energy nuclear experiments, providing guidance for EoS modeling. 
These constraints include the value of $n_0$, binding energy per nucleon, incompressibility, and hyperon optical potentials, among others. An example is the $\Sigma$ potential recently calculated by ALICE \cite{ALICE:2020mfd}, shown to be relevant for NSs (see Fig.~7 of Ref.~\cite{Fabbietti:2020bfg}). Other constraints include the symmetry energy and its derivatives and, of course, NS observables (see Ref.~\cite{Ghosh:2022lam} for a recent review).  The latter apply to zero-temperature $\beta-$equilibrated matter and include reproducing NSs with at least $2$ M$_\odot$, radii and tidal deformability in agreement with new NASA NICER and LIGO/VIRGO results (see Fig.~1 in Ref.~\cite{Tan:2021ahl} for a compilation of data shown in the mass-radius diagram), and cooling data. Chiral effective field theory methods also provide a systematic way to learn about the EoS at low and intermediate densities and temperatures \cite{Hebeler:2013nza,Tews:2018iwm,Lim:2018bkq,Dexheimer:2018dhb}.

Finite-temperature constraints include intermediate and high-energy nuclear experiments and lattice QCD. The (nearly) isospin-symmetric and zero net strangeness heavy-ion constraints require an extrapolation in isospin and strangeness, which is non-trivial at finite temperature, when comparing with NS matter~\cite{Aryal:2020ocm}. The lattice QCD constraints are provided at any isospin and strangeness \cite{Noronha-Hostler:2019ayj}, but are restricted to low density (relative to the temperature) and require a large extrapolation in density when comparing with NS matter. Still, lattice provides an anchor for the QCD phase diagram, determining that deconfinement is a crossover at zero $n_B$ \cite{Aoki:2006we}.

Finally, as either the density, temperature, or both increase, baryons start to overlap, either because they come closer or become larger. At this point, the definition of baryon looses it meaning and the degrees of freedom of high-energy matter become quarks. Although the deconfinement of quark matter inside cold beta-equilibrated NSs is still a open question, it is known from QCD that chiral symmetry should be restored at large densities and temperatures, decreasing the overall baryonic masses and the mass gap between parity doublets \cite{Aarts:2017rrl}, which can have interesting consequences for NSs \cite{Dexheimer:2008cv,Dexheimer:2012eu,Marczenko:2018jui}. At asymptotically large energies, perturbative QCD can provide the weakly interacting EoS \cite{Haque:2013sja,Kurkela:2016was}, which can be used to constrain the NS EoS \cite{Fraga:2013qra,Komoltsev:2021jzg}.

\section{Recent developments}

In the past decade, much discussion has been devoted to possible struc- \clearpage \noindent
ture in the speed of sound $c_s$ of dense matter \cite{Bedaque:2014sqa,Kojo:2009ha,Tews:2018kmu,Baym:2019iky}. This steams from the fact that $2$ M$_\odot$ stars require a stiff EoS (with $c_s^2\to1$ in natural units) at intermediate densities, while the conformal limit of massless free quarks requires $c_s^2\to1/3$ from below at asymptotically large densities. This implies a non-monotonic behavior in $c_s$, usually referred to as a ``bump''. Besides EoS softening due to new degrees of freedom, other explanations for the bump are medium effects related to a vector condensate \cite{Pisarski:2021aoz} or breaking or restoration of symmetries \cite{Hippert:2021gfs,Marczenko:2018jui}. Fig.~1 (adapted from Ref. \cite{Tan:2021ahl}) shows how bumps appear in realistic models and under a controlled $c_s$ parametrization, which allows one to relate the density at which the bump appears with curves in the NS mass-radius diagram. Furthermore, different slopes in the mass-radius diagram manifest as different slopes of the binary Love relation (between the tidal deformabilities of binary NSs), which may be observable during the fifth LIGO observing run (O5) \cite{Tan:2021nat}. See Refs.~\cite{Legred:2022pyp,Altiparmak:2022bke} and references therein for additional discussion on the correlation between the NS EoS and observational constraints.

\begin{figure}[t]
\includegraphics[width=0.325\textwidth]{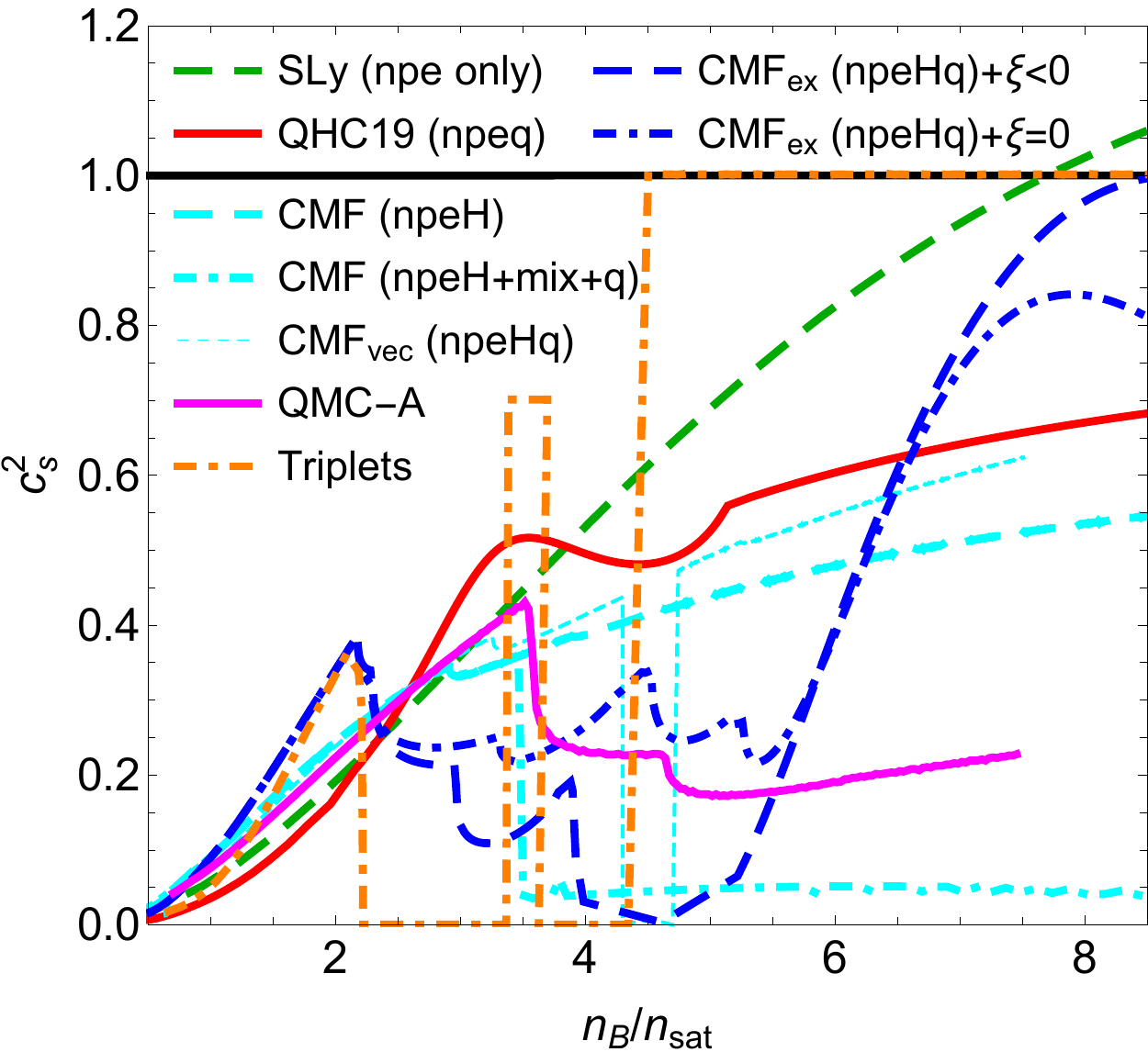}
\includegraphics[width=0.325\textwidth]{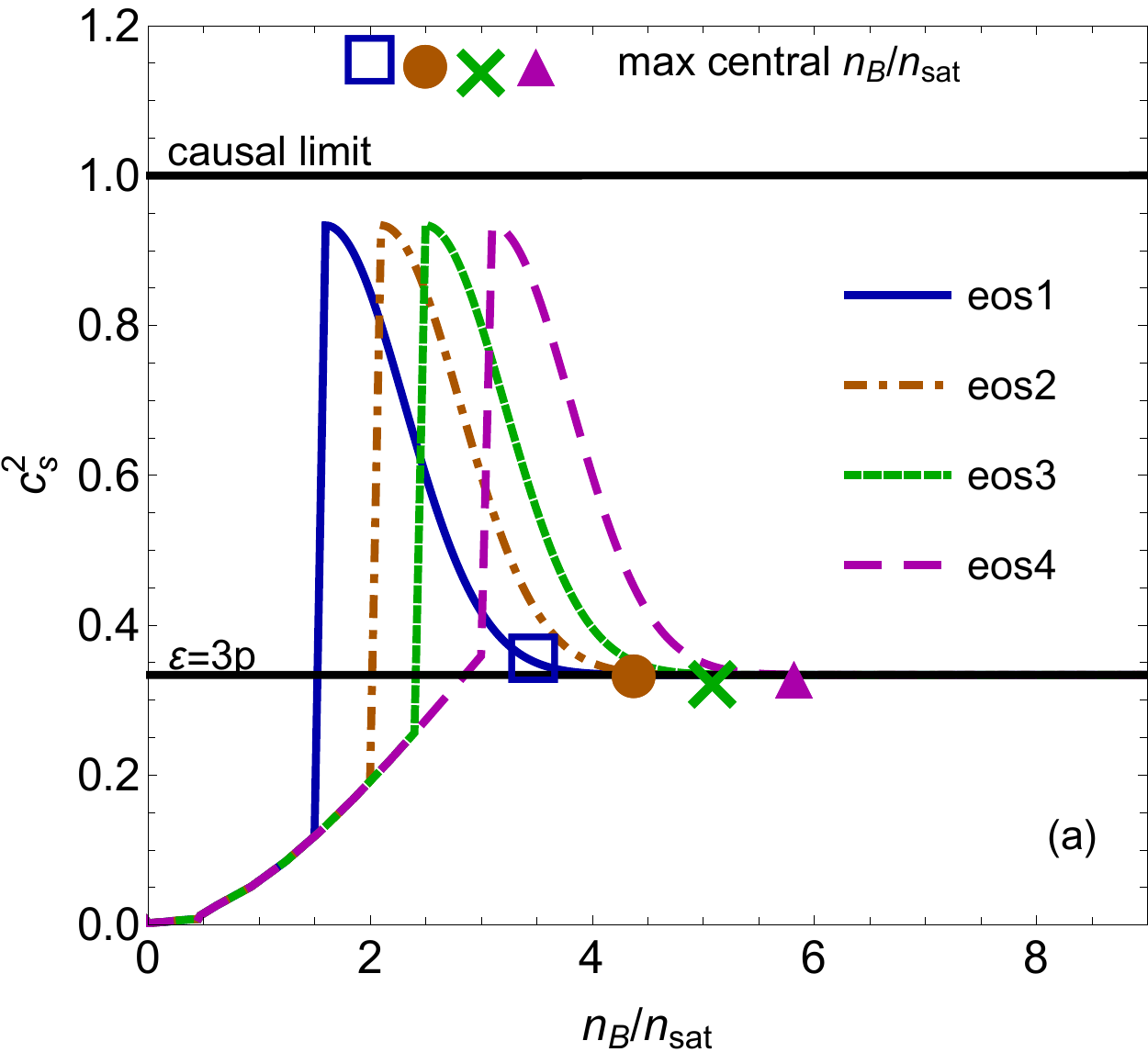}
\includegraphics[width=0.33\textwidth]{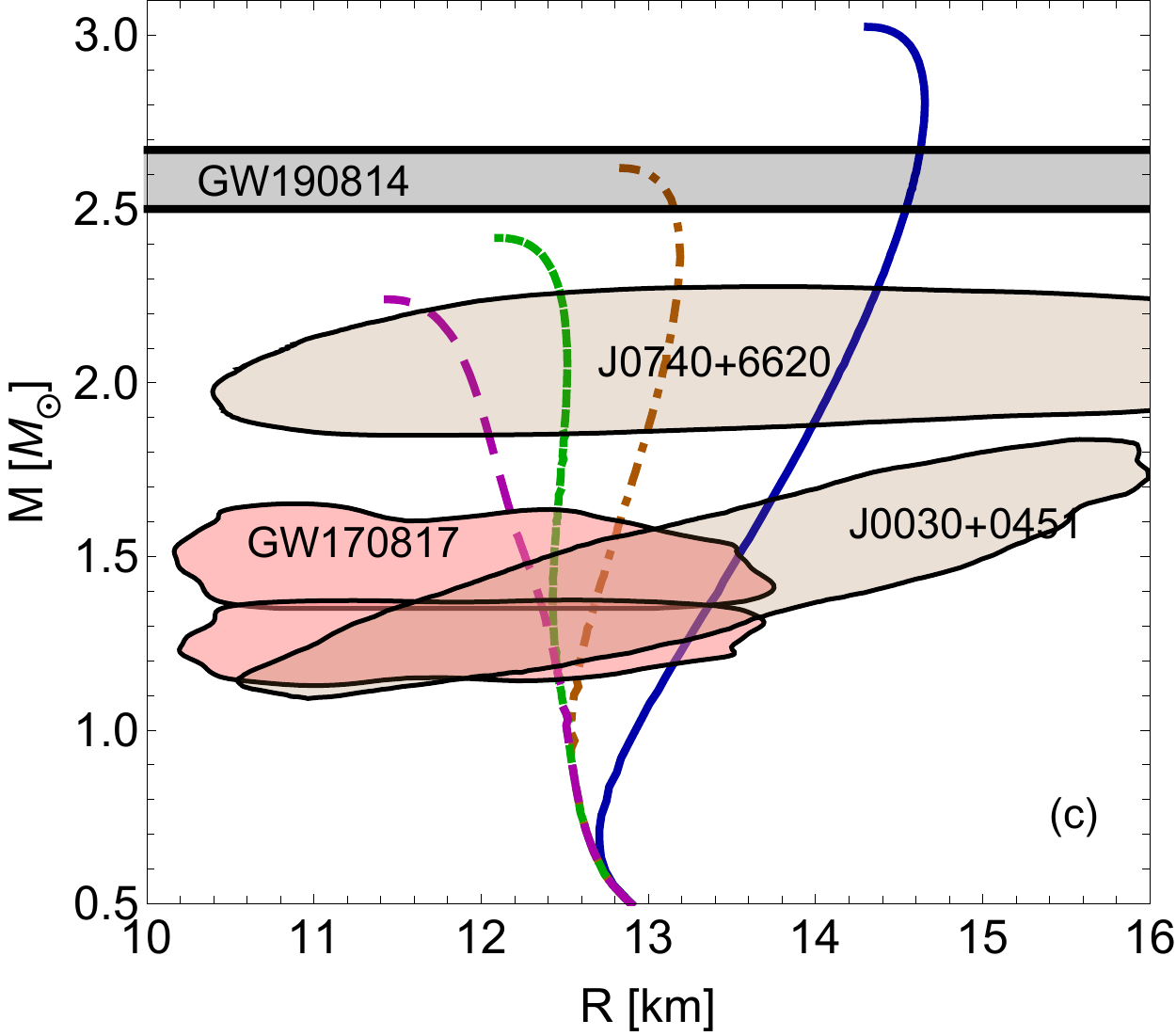}
\caption{Left: speed of sound ``bumps'' generated by several realistic EoSs. \\Middle and Right: relation between density at which speed of sound bumps are present and mass-radius relation of neutron stars within a parametrized approach. Figure modified from Ref.~\cite{Tan:2021ahl}.}
\end{figure}

Beyond NS static properties, signatures for deconfinement to quark matter have been investigated in supernova explosions and in NS mergers. In particular, additional bursts of neutrinos were predicted when a hybrid star is formed already during the supernova stage \cite{Sagert:2008ka,Nakazato:2013iba,Fischer:2017lag,Kuroda:2021eiv,Jakobus:2022ucs}. In NS mergers, signals for deconfinement as a first-order phase transition have been identified in the waveform, through changes in amplitude, frequency, and duration of the postmerger signal \cite{Most:2018eaw,Aloy:2018jov,Weih:2019xvw}, which has not yet been detected. Through the identification of an universal relation including the postmerger peak frequency and tidal deformability, a deviation from it was identified as a signature for deconfinement \cite{Blacker:2020nlq}. There were also other effects related to smoothing out the first-order phase transition using piecewise polytropes~\cite{Fujimoto:2022xhv}, a percolation scenario \cite{Huang:2022mqp}, the construction of a mixture of phases (Gibbs construction) \cite{Prakash:2021wpz}, or a crossover \cite{Most:2022wgo} that were identified in merger simulations. In all these studies, the temperature dependence was included in the EoS, either self-consistently, which allows different degrees of freedom and the phase transition itself to depend on temperature (see Refs.~\cite{Raduta:2021coc,Kanakis-Pegios:2022abm,Raduta:2022elz} for details), or in an ad hoc manner. 

\section{Comparison of neutron-star mergers and heavy-ion collisions}

\begin{figure}[t]
\centering
\includegraphics[width=0.71\textwidth,trim={0 0 0 0.15cm},clip]{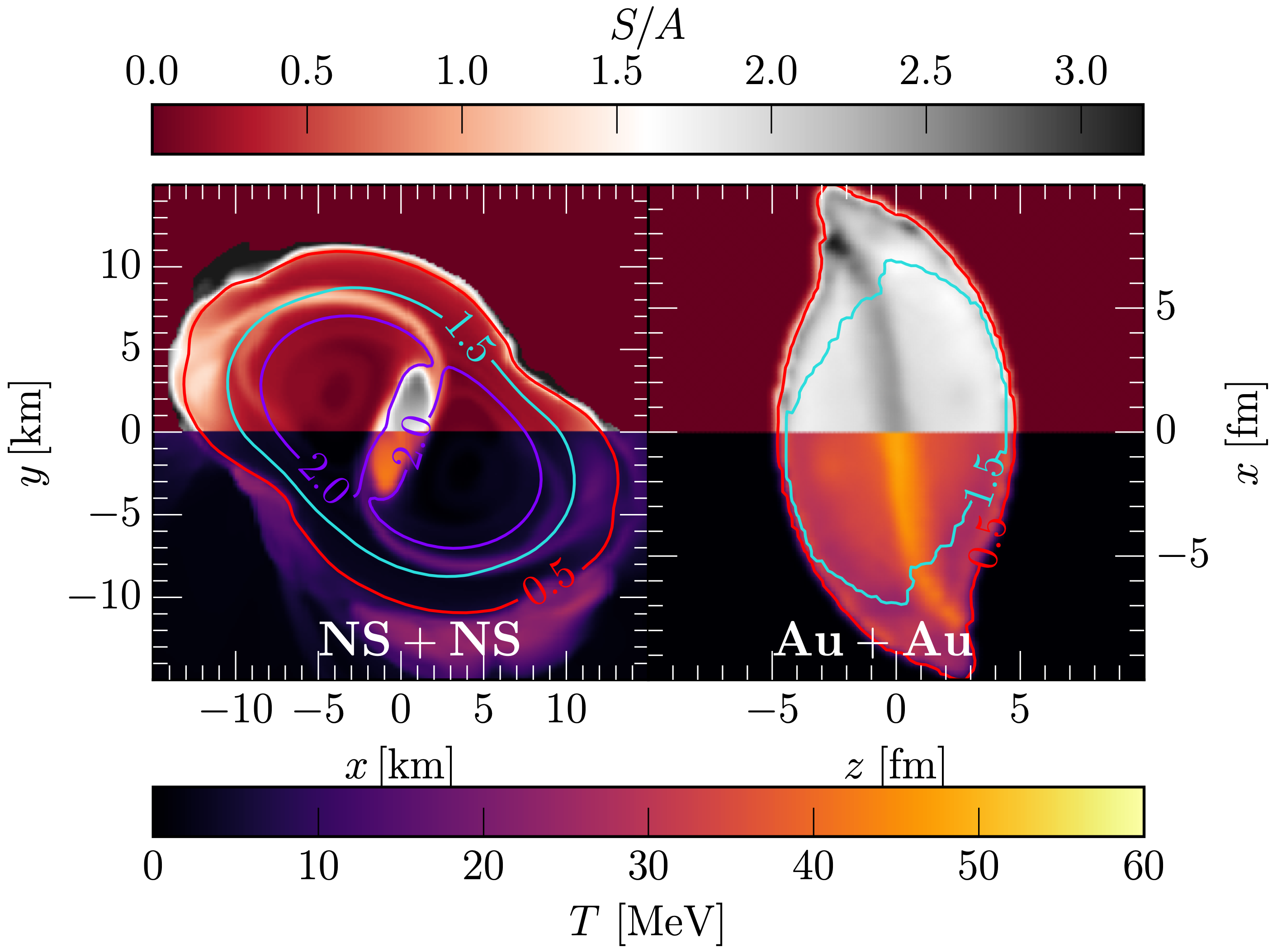}
\caption{ Distributions of entropy per baryon (top) and temperature (bottom) for a binary neutron-star merger
with total mass of $2.8$ M$_\odot$ (left) and a Au+Au heavy-ion collision at E$_{\rm{lab}}=450$ MeV (right). Density contours are shown in units of $n_0$. The snapshots refer
to $t = 3$ ms after merger and to
$t = 5$ fm/c after the full collision overlap. Figure modified from Ref.~\cite{Most:2022wgo}.}
\end{figure}

Recently, the first self-consistent comparison between NS mergers and low-energy heavy-ion collisions was presented in Ref.~\cite{Most:2022wgo}. It used the SU(3) CMF model with excluded volume including parity doublets \cite{Mukherjee:2017jzi,Motornenko:2019arp} in relativistic hydrodynamics simulations using the full general relativity Frankfurt/Illinois GRMHD code \cite{Etienne_2015} and the Frankfurt SHASTA code. Fig.~2 shows how, in spite of a difference of 18 orders of magnitude between the systems, their geometry is very similar. It was shown that, despite some differences generated by gravity (which is only relevant for the merger), similar temperatures, entropies and, most importantly, densities are achieved. In particular, similar trajectories in the QCD phase diagram allow for the first time to establish a connection between different total merger masses and laboratory energies, presently available for experiments of the HADES detector set-up at GSI \cite{HADES:2019auv}. This kind of comparison is finally allowing us to quantitatively study astrophysics in the laboratory.

\section*{Acknowledgements}

Support for this work comes from the National Science Foundation under grants PHY1748621, MUSES OAC-2103680, and NP3M PHY-2116686. 

\bibliographystyle{IEEEtran}  
\bibliography{bibliography} 

\begin{thebibliography}{10}
\providecommand{\url}[1]{#1}
\csname url@samestyle\endcsname
\providecommand{\newblock}{\relax}
\providecommand{\bibinfo}[2]{#2}
\providecommand{\BIBentrySTDinterwordspacing}{\spaceskip=0pt\relax}
\providecommand{\BIBentryALTinterwordstretchfactor}{4}
\providecommand{\BIBentryALTinterwordspacing}{\spaceskip=\fontdimen2\font plus
\BIBentryALTinterwordstretchfactor\fontdimen3\font minus
  \fontdimen4\font\relax}
\providecommand{\BIBforeignlanguage}[2]{{%
\expandafter\ifx\csname l@#1\endcsname\relax
\typeout{** WARNING: IEEEtran.bst: No hyphenation pattern has been}%
\typeout{** loaded for the language `#1'. Using the pattern for}%
\typeout{** the default language instead.}%
\else
\language=\csname l@#1\endcsname
\fi
#2}}
\providecommand{\BIBdecl}{\relax}
\BIBdecl

\bibitem{Sagert:2005fw}
I.~Sagert, M.~Hempel, C.~Greiner, and J.~Schaffner-Bielich, \emph{Eur. J.
  Phys.}, vol.~27, pp. 577--610, 2006.

\bibitem{Glendenning:1982nc}
N.~K. Glendenning, \emph{Phys. Lett. B}, vol. 114, pp. 392--396, 1982.

\bibitem{Prakash:1992zng}
M.~Prakash, M.~Prakash, J.~M. Lattimer, and C.~J. Pethick, \emph{Astrophys. J.
  Lett.}, vol. 390, p. L77, 1992.

\bibitem{Weber:2014qoa}
F.~Weber, G.~A. Contrera, M.~G. Orsaria, W.~Spinella, and O.~Zubairi,
  \emph{Mod. Phys. Lett. A}, vol.~29, p. 1430022, 2014.

\bibitem{Horowitz:2004yf}
C.~J. Horowitz, M.~A. Perez-Garcia, and J.~Piekarewicz, \emph{Phys. Rev. C},
  vol.~69, p. 045804, 2004.

\bibitem{Baldo:1992kzz}
M.~Baldo, J.~Cugnon, A.~Lejeune, and U.~Lombardo, \emph{Nucl. Phys. A}, vol.
  536, pp. 349--365, 1992.

\bibitem{Page:2010aw}
D.~Page, M.~Prakash, J.~M. Lattimer, and A.~W. Steiner, \emph{Phys. Rev.
  Lett.}, vol. 106, p. 081101, 2011.

\bibitem{Sedrakian:2004yq}
A.~Sedrakian, \emph{Phys. Rev. D}, vol.~71, p. 083003, 2005.

\bibitem{Alford:2001dt}
M.~G. Alford, \emph{Ann. Rev. Nucl. Part. Sci.}, vol.~51, pp. 131--160, 2001.

\bibitem{Ivanytskyi:2022oxv}
O.~Ivanytskyi and D.~Blaschke, arXiv: 2204.03611 [nucl-th] 2022.

\bibitem{compose}
\BIBentryALTinterwordspacing
 [Online]. Available: \url{https://compose.obspm.fr/}
\BIBentrySTDinterwordspacing

\bibitem{Typel:2022lcx}
S.~Typel \emph{et~al.}, arXiv: 2203.03209 [astro-ph.HE] 2022.

\bibitem{muses}
\BIBentryALTinterwordspacing
 [Online]. Available: \url{https://muses.physics.illinois.edu/}
\BIBentrySTDinterwordspacing

\bibitem{Zeldovich:1961sbr}
Y.~B. Zel'dovich, \emph{Zh. Eksp. Teor. Fiz.}, vol.~41, pp. 1609--1615, 1961.

\bibitem{ALICE:2020mfd}
S.~Acharya \emph{et~al.}, \emph{Nature}, vol. 588, pp. 232--238, 2020,
  [Erratum: Nature 590, E13 (2021)].

\bibitem{Fabbietti:2020bfg}
L.~Fabbietti, V.~Mantovani~Sarti, and O.~Vazquez~Doce, \emph{Ann. Rev. Nucl.
  Part. Sci.}, vol.~71, pp. 377--402, 2021.

\bibitem{Ghosh:2022lam}
S.~Ghosh, B.~K. Pradhan, D.~Chatterjee, and J.~Schaffner-Bielich, \emph{Front.
  Astron. Space Sci.}, vol.~9, p. 864294, 2022.

\bibitem{Tan:2021ahl}
H.~Tan, T.~Dore, V.~Dexheimer, J.~Noronha-Hostler, and N.~Yunes, \emph{Phys.
  Rev. D}, vol. 105, no.~2, p. 023018, 2022.

\bibitem{Hebeler:2013nza}
K.~Hebeler, J.~M. Lattimer, C.~J. Pethick, and A.~Schwenk, \emph{Astrophys.
  J.}, vol. 773, p.~11, 2013.

\bibitem{Tews:2018iwm}
I.~Tews, J.~Margueron, and S.~Reddy, \emph{Phys. Rev. C}, vol.~98, no.~4, p.
  045804, 2018.

\bibitem{Lim:2018bkq}
Y.~Lim and J.~W. Holt, \emph{Phys. Rev. Lett.}, vol. 121, no.~6, p. 062701,
  2018.

\bibitem{Dexheimer:2018dhb}
V.~Dexheimer, R.~de~Oliveira~Gomes, S.~Schramm, and H.~Pais, \emph{J. Phys. G},
  vol.~46, no.~3, p. 034002, 2019.

\bibitem{Aryal:2020ocm}
K.~Aryal, C.~Constantinou, R.~L.~S. Farias, and V.~Dexheimer, \emph{Phys. Rev.
  D}, vol. 102, no.~7, p. 076016, 2020.

\bibitem{Noronha-Hostler:2019ayj}
J.~Noronha-Hostler, P.~Parotto, C.~Ratti, and J.~M. Stafford, \emph{Phys. Rev.
  C}, vol. 100, no.~6, p. 064910, 2019.

\bibitem{Aoki:2006we}
Y.~Aoki, G.~Endrodi, Z.~Fodor, S.~D. Katz, and K.~K. Szabo, \emph{Nature}, vol.
  443, pp. 675--678, 2006.

\bibitem{Aarts:2017rrl}
G.~Aarts, C.~Allton, D.~De~Boni, S.~Hands, B.~J\"ager, C.~Praki, and J.-I.
  Skullerud, \emph{JHEP}, vol.~06, p. 034, 2017.

\bibitem{Dexheimer:2008cv}
V.~Dexheimer, G.~Pagliara, L.~Tolos, J.~Schaffner-Bielich, and S.~Schramm,
  \emph{Eur. Phys. J. A}, vol.~38, pp. 105--113, 2008.

\bibitem{Dexheimer:2012eu}
V.~Dexheimer, J.~Steinheimer, R.~Negreiros, and S.~Schramm, \emph{Phys. Rev.
  C}, vol.~87, no.~1, p. 015804, 2013.

\bibitem{Marczenko:2018jui}
M.~Marczenko, D.~Blaschke, K.~Redlich, and C.~Sasaki, \emph{Phys. Rev. D},
  vol.~98, no.~10, p. 103021, 2018.

\bibitem{Haque:2013sja}
N.~Haque, J.~O. Andersen, M.~G. Mustafa, M.~Strickland, and N.~Su, \emph{Phys.
  Rev. D}, vol.~89, no.~6, p. 061701, 2014.

\bibitem{Kurkela:2016was}
A.~Kurkela and A.~Vuorinen, \emph{Phys. Rev. Lett.}, vol. 117, no.~4, p.
  042501, 2016.

\bibitem{Fraga:2013qra}
E.~S. Fraga, A.~Kurkela, and A.~Vuorinen, \emph{Astrophys. J. Lett.}, vol. 781,
  no.~2, p. L25, 2014.

\bibitem{Komoltsev:2021jzg}
O.~Komoltsev and A.~Kurkela, \emph{Phys. Rev. Lett.}, vol. 128, no.~20, p.
  202701, 2022.

\bibitem{Bedaque:2014sqa}
P.~Bedaque and A.~W. Steiner, \emph{Phys. Rev. Lett.}, vol. 114, no.~3, p.
  031103, 2015.

\bibitem{Kojo:2009ha}
T.~Kojo, Y.~Hidaka, L.~McLerran, and R.~D. Pisarski, \emph{Nucl. Phys. A}, vol.
  843, pp. 37--58, 2010.

\bibitem{Tews:2018kmu}
I.~Tews, J.~Carlson, S.~Gandolfi, and S.~Reddy, \emph{Astrophys. J.}, vol. 860,
  no.~2, p. 149, 2018.

\bibitem{Baym:2019iky}
G.~Baym, S.~Furusawa, T.~Hatsuda, T.~Kojo, and H.~Togashi, \emph{Astrophys.
  J.}, vol. 885, p.~42, 2019.

\bibitem{Pisarski:2021aoz}
R.~D. Pisarski, \emph{Phys. Rev. D}, vol. 103, no.~7, p. L071504, 2021.

\bibitem{Hippert:2021gfs}
M.~Hippert, E.~S. Fraga, and J.~Noronha, \emph{Phys. Rev. D}, vol. 104, no.~3,
  p. 034011, 2021.

\bibitem{Tan:2021nat}
H.~Tan, V.~Dexheimer, J.~Noronha-Hostler, and N.~Yunes, \emph{Phys. Rev.
  Lett.}, vol. 128, no.~16, p. 161101, 2022.

\bibitem{Legred:2022pyp}
I.~Legred, K.~Chatziioannou, R.~Essick, and P.~Landry, \emph{Phys. Rev. D},
  vol. 105, no.~4, p. 043016, 2022.

\bibitem{Altiparmak:2022bke}
S.~Altiparmak, C.~Ecker, and L.~Rezzolla, arXiv: 2203.14974 [astro-ph.HE] 2022.

\bibitem{Sagert:2008ka}
I.~Sagert, T.~Fischer, M.~Hempel, G.~Pagliara, J.~Schaffner-Bielich,
  A.~Mezzacappa, F.~K. Thielemann, and M.~Liebendorfer, \emph{Phys. Rev.
  Lett.}, vol. 102, p. 081101, 2009.

\bibitem{Nakazato:2013iba}
K.~Nakazato, K.~Sumiyoshi, and S.~Yamada, \emph{Astron. Astrophys.}, vol. 558,
  p. A50, 2013.

\bibitem{Fischer:2017lag}
T.~Fischer, N.-U.~F. Bastian, M.-R. Wu, P.~Baklanov, E.~Sorokina, S.~Blinnikov,
  S.~Typel, T.~Kl\"ahn, and D.~B. Blaschke, \emph{Nature Astron.}, vol.~2,
  no.~12, pp. 980--986, 2018.

\bibitem{Kuroda:2021eiv}
T.~Kuroda, T.~Fischer, T.~Takiwaki, and K.~Kotake, \emph{Astrophys. J.}, vol.
  924, no.~1, p.~38, 2022.

\bibitem{Jakobus:2022ucs}
P.~Jakobus, B.~Mueller, A.~Heger, A.~Motornenko, J.~Steinheimer, and
  H.~Stoecker, arXiv: 2204.10397 [astro-ph.HE] 2022.

\bibitem{Most:2018eaw}
E.~R. Most, L.~J. Papenfort, V.~Dexheimer, M.~Hanauske, S.~Schramm,
  H.~St\"ocker, and L.~Rezzolla, \emph{Phys. Rev. Lett.}, vol. 122, no.~6, p.
  061101, 2019.

\bibitem{Aloy:2018jov}
M.~A. Aloy, J.~M. Ib\'a\~nez, N.~Sanchis-Gual, M.~Obergaulinger, J.~A. Font,
  S.~Serna, and A.~Marquina, \emph{Mon. Not. Roy. Astron. Soc.}, vol. 484, p.
  4980, 2019.

\bibitem{Weih:2019xvw}
L.~R. Weih, M.~Hanauske, and L.~Rezzolla, \emph{Phys. Rev. Lett.}, vol. 124,
  no.~17, p. 171103, 2020.

\bibitem{Blacker:2020nlq}
S.~Blacker, N.-U.~F. Bastian, A.~Bauswein, D.~B. Blaschke, T.~Fischer,
  M.~Oertel, T.~Soultanis, and S.~Typel, \emph{Phys. Rev. D}, vol. 102, no.~12,
  p. 123023, 2020.

\bibitem{Fujimoto:2022xhv}
Y.~Fujimoto, K.~Fukushima, K.~Hotokezaka, and K.~Kyutoku, arXiv: 2205.03882
  [astro-ph.HE] 2022.

\bibitem{Huang:2022mqp}
Y.-J. Huang, L.~Baiotti, T.~Kojo, K.~Takami, H.~Sotani, H.~Togashi, T.~Hatsuda,
  S.~Nagataki, and Y.-Z. Fan, arXiv: 2203.04528 [astro-ph.HE] 2022.

\bibitem{Prakash:2021wpz}
A.~Prakash, D.~Radice, D.~Logoteta, A.~Perego, V.~Nedora, I.~Bombaci,
  R.~Kashyap, S.~Bernuzzi, and A.~Endrizzi, \emph{Phys. Rev. D}, vol. 104,
  no.~8, p. 083029, 2021.

\bibitem{Most:2022wgo}
E.~R. Most, A.~Motornenko, J.~Steinheimer, V.~Dexheimer, M.~Hanauske,
  L.~Rezzolla, and H.~Stoecker, arXiv: 2201.13150 [nucl-th] 2022.

\bibitem{Raduta:2021coc}
A.~R. Raduta, F.~Nacu, and M.~Oertel, \emph{Eur. Phys. J. A}, vol.~57, no.~12,
  p. 329, 2021.

\bibitem{Kanakis-Pegios:2022abm}
A.~Kanakis-Pegios, P.~S. Koliogiannis, and C.~C. Moustakidis, arXiv: 2202.01820
  [astro-ph.HE] 2022.

\bibitem{Raduta:2022elz}
A.~R. Raduta, arXiv: 2205.03177 [nucl-th] 2022.

\bibitem{Mukherjee:2017jzi}
A.~Mukherjee, S.~Schramm, J.~Steinheimer, and V.~Dexheimer, \emph{Astron.
  Astrophys.}, vol. 608, p. A110, 2017.

\bibitem{Motornenko:2019arp}
A.~Motornenko, J.~Steinheimer, V.~Vovchenko, S.~Schramm, and H.~Stoecker,
  \emph{Phys. Rev. C}, vol. 101, no.~3, p. 034904, 2020.

\bibitem{Etienne_2015}
Z.~B. Etienne, V.~Paschalidis, R.~Haas, P.~Mösta, and S.~L. Shapiro,
  \emph{Classical and Quantum Gravity}, vol.~32, no.~17, p. 175009, aug 2015.

\bibitem{HADES:2019auv}
J.~Adamczewski-Musch \emph{et~al.}, \emph{Nature Phys.}, vol.~15, no.~10, pp.
  1040--1045, 2019.

\end{thebibliography}

\end{document}